\documentclass[aps,prl,reprint,showpacs,citeautoscript,superscriptaddress
]{revtex4-1} 

\usepackage{graphicx} 
\usepackage{dcolumn}  
\usepackage{bm}       
\usepackage{amsmath}  
\usepackage{color}    

\begin{document}

\title{The electron elevator: excitations across the band gap via a
  dynamical gap state}

\author{A. Lim}
\email{anthony.lim06@imperial.ac.uk}
\affiliation{Department of Physics and Thomas Young Centre, Imperial
  College London, London SW7 2AZ, United Kingdom}%

\author{W. M. C. Foulkes}%
\affiliation{Department of Physics and Thomas Young Centre, Imperial
  College London, London SW7 2AZ, United Kingdom}%

\author{A. P. Horsfield}
\affiliation{Department of Materials and Thomas Young Centre, Imperial
  College London, London SW7 2AZ, United Kingdom}%

\author{D. R. Mason}
\affiliation{CCFE, Culham Centre for Fusion Energy, Abingdon, Oxfordshire OX14 3DB, United Kingdom }%

\author{A. Schleife}
\affiliation{Lawrence Livermore National Laboratory, 7000 East Avenue,
  Livermore, CA 94550, USA}%

\author{E. W. Draeger}
\affiliation{Lawrence Livermore National Laboratory, 7000 East Avenue,
  Livermore, CA 94550, USA}%

\author{A. A. Correa}
\affiliation{Lawrence Livermore National Laboratory, 7000 East Avenue,
  Livermore, CA 94550, USA}%

\date{\today}

\begin{abstract}

We have used time-dependent density functional theory to study self-irradiated Si.
We calculate the electronic stopping power of Si in Si by evaluating the energy transferred to the electrons per unit path length by an ion of kinetic energy from $1\,\text{eV}$ to $100\,\text{keV}$ moving through the host.
Electronic stopping is found to be significant below the threshold velocity normally identified with direct transitions across the band gap.
A structured crossover at low velocity exists in place of a hard threshold.
An analysis of the time dependence of the transition rates using coupled linear rate equations enables one of the excitation mechanisms to be clearly identified: a defect state induced in the gap by the moving ion acts like an elevator and carries electrons across the band gap.

\end{abstract}

\pacs{Valid PACS appear here}

\maketitle

To maintain useful operational lifetimes of nuclear facilities and space equipment, their components need to be resilient to radiation \cite{NASA,stott_page_276, fusion}.
For this reason, the effects of radiation have been extensively studied, experimentally and theoretically.
Most of the scientific instruments (e.g., detectors) used in the nuclear and space industries are made from semiconductors that may be damaged when exposed to radiation, so it is of critical importance to understand how radiation damage in semiconductors is initiated and evolves over time.

Our understanding of the effects of radiation on materials is largely based on a classical picture of ion-ion collisions.
Only recently have improvements in quantum mechanical electronic structure techniques and computational facilities allowed the electronic contributions to be investigated quantitatively \cite{Au,LiF, Al, mason_chnl, mason_MD, mason_atomic_dyn}.
The stopping power --- the kinetic energy lost per unit path length by a particle (projectile) moving through a target (host) material \footnote{Note that the quantity conventionally known as the stopping power has the dimensions of a force.}  --- has historically been divided into ionic and electronic components \cite{SRIM}.
Electronic stopping is most important at high velocities or under channeling conditions, when the projectile travels a large distance without undergoing direct collisions with host atoms.
A projectile on a channeling trajectory mainly perturbs the electron density.

The material studied in this Letter is self-irradiated silicon at low projectile velocities (such as might be produced, e.g., by a primary knock-on ion or secondary events).
The projectile kinetic energy $K$ ranges from $1\,\text{eV}$ to $100\,\text{keV}$, which is low enough to ensure that the evolution of the electronic states is only weakly non-adiabatic.

It is the behavior of the cascade that dictates the final distribution of damage \cite{duffy}.
Semiconductors and insulators amorphize on irradiation \cite{trachenko2004understanding}, with the degree of amorphization largely governed by the interplay between short-range covalent, and long-range Coulombic forces \cite{trachenko2005nature}.
Electronic stopping of high-energy atoms reduces the peak size of the damaged region \cite{zarkadoula2014high}, as well as the extent of the residual damage after partial recrystallization \cite{phillips2010two}.
The electronic stopping of lower energy atoms in insulators is less well understood but could be equally significant: in metals, electronic stopping of low-energy atoms can dominate energy transfer, as there are so many of them during the cooling phase of a cascade \cite{duvenbeck2005low,sand2015lower,mason2015incorporating}.


Details of the electronic structure (such as the band gap and the density of states) play a fundamental role at low projectile velocities, so we use a first-principles molecular dynamics technique in which the electronic excitations are described by time-dependent density functional theory (TDDFT) with no adjustable parameters.
This approach captures the complicated band-structure features of the silicon-plus-silicon system without treating the projectile as a weak perturbation.
TDDFT complements a variety of other atomistic and non-atomistic techniques with adjustable parameters \cite{mason_chnl, nagy, emilio}.
Based on our simulation results and a quantum mechanical band-structure picture, we propose a coupled-rate-equation model that explains the electronic stopping mechanisms in this regime.

It has been suggested that there is a threshold velocity below which a light projectile (such as a proton) moving through a wide-band-gap insulator (such as LiF) is unable to exchange sufficient energy to excite electrons across the band gap \cite{LiF,LiF_SiO2}.
The electronic stopping would be strictly zero below this threshold.
An estimate of the threshold velocity follows from the observation that a projectile channeling through a crystal experiences a time-varying potential oscillating at the atom-passing frequency, $f=v/\lambda$, where $\lambda$ is the distance between equivalent lattice positions.
(For the $\langle 001 \rangle$ channel of Si, $\lambda = a/4$, where $a$ is the lattice parameter.)
The threshold velocity, $v_\text{th}$, may be obtained by equating $hf$ ($h$ is Planck's constant) to the band gap $\Delta$, giving
\begin{equation} 
v_\text{th} = \frac{\lambda\Delta}{h} .  \label{atompass} 
\end{equation}
An alternative estimate may be obtained by evaluating the energy and momentum needed to create the lowest-energy electron-hole excitation in a homogeneous electron gas with an energy gap $\Delta$.
This yields \cite{levine_louie}
\begin{equation}
v_\text{th} = \frac{\Delta}{2\hbar
  k_\text{F}}=\left(\frac{2\pi}{3}\right)^\frac{1}{3}
\frac{\lambda\Delta}{h} , \label{di}
\end{equation}
where $k_\text{F}$ is the Fermi vector of an effective jellium with electron density equal to the average for the insulator.
The first equality may be derived from the premise that an ion must lose energy greater than $\Delta$ to excite an electron across the gap.
For a non-relativistic ion of kinetic energy $m v^2/2 \gg \Delta$, this implies $|\bm{v}\cdot\Delta \bm{p}| > \Delta$, where $\Delta \bm{p}$ is the momentum transferred to the electron.
Since the ion is slow and massive, $|\Delta \bm{p}| \lesssim 2\hbar k_F$ and the first equality follows.
The second equality is obtained by expressing $k_F$ in terms of the number density of valence electrons in Si and assuming that the projectile is moving down a $\langle 001 \rangle$ channel.
In practice, the thresholds given by Eqs.~\ref{atompass} and \ref{di} are very similar despite their different physical origins.
For a material with the valence electron density of silicon and a gap of $0.6\,\text{eV}$ (the band gap of Si according to DFT within the local density approximation), the thresholds are $0.20$ and $0.25\,\text{\AA\,fs}^{-1}$ ($K = 56$ and $92$\,eV).

Artacho \cite{emilio} approached the problem of electronic stopping using a simple 1D model with time-dependent hopping between flat bands.
He showed that the dependence of the stopping on velocity can be quite intricate and that, although the stopping vanishes as $v\to0$, there is no hard threshold.

Experimentally, the presence of a threshold is debatable \cite{auth, markin, eder} due to difficulties in separating the nuclear and electronic stopping powers at low velocity \cite{zhang}.
Auth \textit{et al.}\ \cite{auth} and Markin \textit{et al.}\ \cite{markin} considered channeling ions in a LiF crystal and both claimed the presence of a hard threshold.
However, Eder \textit{et al.}\ \cite{eder} did not observe a threshold for protons in LiF, arguing that below-threshold stopping is possible due to a process called molecular orbital promotion, in which the local band gap is reduced as the channeling ion passes through the crystal.
 Motivated by these open questions in experiment and theory, the simulation results reported in this Letter suggest a simple resolution of the velocity threshold debate.

Our TDDFT simulations were carried out using a non-adiabatic modification \cite{qbox} of the first-principles quantum molecular dynamics code \textsc{Qbox} \cite{gygi} to investigate a Si in Si system.
The adiabatic local density approximation (ALDA) to exchange and correlation was used throughout.
The periodic supercell contained $216$ stationary Si atoms in the diamond structure plus the channeling Si atom (a total of $868$ valence electrons).
The Si projectiles of most interest in this work have kinetic energies of less than $1 \text{keV}$, which is too low for any significant channeling to take place.
However, the interpretation of the results is much simplified by dragging the projectile at constant velocity along a channel in an otherwise stationary crystal, allowing electrons to become excited according to the time-dependent Kohn-Sham equations \cite{rungegross}.
We discuss the relationship between these idealized simulations and reality later on.
The wavefunctions were expanded in a plane-wave basis with an energy cut-off of $680\,\text{eV}$; the supercell $\Gamma$ point was used for k-point sampling.
The wavefunctions were evolved in time using the fourth-order Runge-Kutta algorithm \cite{qbox} and a time step of $0.00242$ fs.
Halving the timestep did not significantly alter the results.

To obtain the excitation energy $\Delta E(z)$ of the electrons at a given position $z$ of the channeling ion, we subtracted the Born-Oppenheimer energy (the ground-state energy at the current position of the channeling ion) from the non-adiabatic energy (the time-dependent Kohn-Sham energy \cite{qbox} of the system including the channeling ion).
The calculation was repeated for each projectile velocity~\cite{LiF}.

\begin{figure}
\includegraphics[trim = 0 0 0 0, clip, scale=0.3]{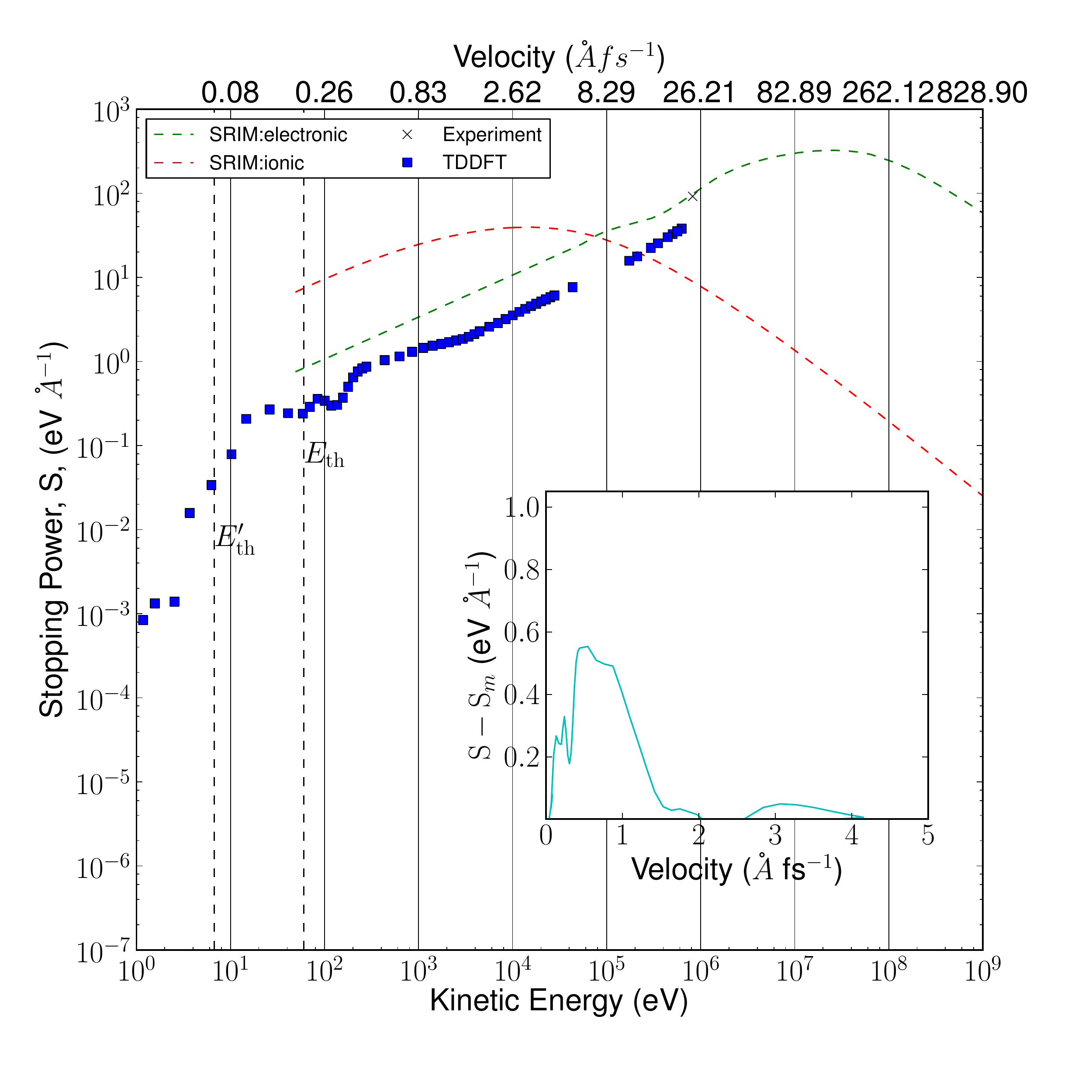}
\caption{ 
Electronic stopping power of a Si self-interstitial moving along the $\langle 001 \rangle$ channel as a function of projectile kinetic energy $K$.
$E_\text{th}$ and $E'_\text{th}$ are calculated using Eq.~(\ref{atompass}) with the LDA band gap and an energy difference of $0.2\,$eV, respectively.
There are three distinct regions: in the metal-like regime ($K > 3000\,\text{eV}$, $v>1.45\,\text{\AA\,fs}^{-1}$) the stopping is linear in projectile velocity, with a slope of $1.47\,\text{eV\,fs\,\AA}^{-2}$; in the band-edge regime ($60 < K < 3000\,\text{eV}$, $0.2<v<1.45\,\text{\AA\,fs}^{-1}$) the stopping rises rapidly with velocity; and in the pre-threshold regime ($K < 60\,\text{eV}$, $v<0.2\,\text{\AA\,fs}^{-1}$) the stopping is non-zero.
The insert shows the stopping power as a function of ion velocity minus the metallic stopping power (defined as $S_m = \gamma\times(v - v_0)$ for $v > v_0$, where $\gamma$ and $v_0$ are fitted to the high velocity points).
Stopping powers from the \textsc{Srim} model \cite{SRIM} (extrapolated to low velocity) and experiment (lowest velocity point measured in Ref.~\cite{zhang}) are also shown.
}\label{stop_e}
\end{figure}

Figure~\ref{stop_e} shows the simulated electronic stopping for motion along the $\langle 001 \rangle$ channel.
The stopping power has three distinct regions: (i) a high-energy metal-like regime where the stopping varies linearly with velocity; (ii) an intermediate (band-edge) regime where there is a rapid rise in the stopping power with velocity; and (iii) a pre-threshold regime at very low velocities.
We note that a non-zero electronic stopping power is observed at velocities well below the thresholds estimated above.

Our calculated stopping power is about a factor of two smaller than predicted by the \textsc{Srim} model \cite{SRIM}, which is fitted to experiment at higher velocities and extrapolated to low velocities.
This may be because channeling ions in experiments do not travel exactly along the center of a silicon channel \cite{zhang}.
Off-center channeling has been reported to increase the electronic stopping power by approximately a factor of two for several materials \cite{dorado,Schleife2015}.
Head-on collisions with lattice atoms also increase the electronic stopping.

At kinetic energies greater than $3000\,\text{eV}$ ($v$$=$$1.45\,\text{\AA\,fs}^{-1}$), the electronic stopping power is directly proportional to the channeling velocity, as shown by Fig.~\ref{stop_e}.
At these high velocities, electronic energy transfer is dominated by direct excitations from deep in the valence band to high in the conduction band and the band gap becomes irrelevant.
The DFT band-gap error is also irrelevant.
We refer to this as the metallic stopping regime.

According to Eqs.~(\ref{atompass}) and $(\ref{di})$, a Si projectile in Si needs $K \gtrsim 60\,\text{eV}$ to excite electrons across the LDA band gap.
The true gap is larger than the LDA gap, so the true kinetic energy threshold is almost certainly larger than $60\,\text{eV}$, but this quantitative error is unlikely to affect the qualitative nature of our results or our conclusions.
Figure \ref{stop_e} shows a rapid increase in stopping for projectile energies between $60$ and $3000\,$eV.
We believe this is due to the rapid increase in the number of states available for direct transitions as the projectile kinetic energy increases and states deeper in the valence and conduction bands become accessible.
Hence, we refer to this as the band-edge regime.

In addition, Fig.~\ref{stop_e} shows a structured non-zero electronic stopping power in the pre-threshold regime below $60\,\text{eV}$ ($0.2\,\text{\AA\,fs}^{-1}$).
This region is important for the later stages of a collision cascade when the structure of the permanent damage is determined \cite{nordlund, duffy}.
Further simulations using Tight-Binding (not reported here) have shown analogous sub-threshold stopping for other projectiles in Si, and we would expect qualitatively similar results to be obtained for other host semiconductors.

\begin{figure}
\includegraphics[scale=0.45]{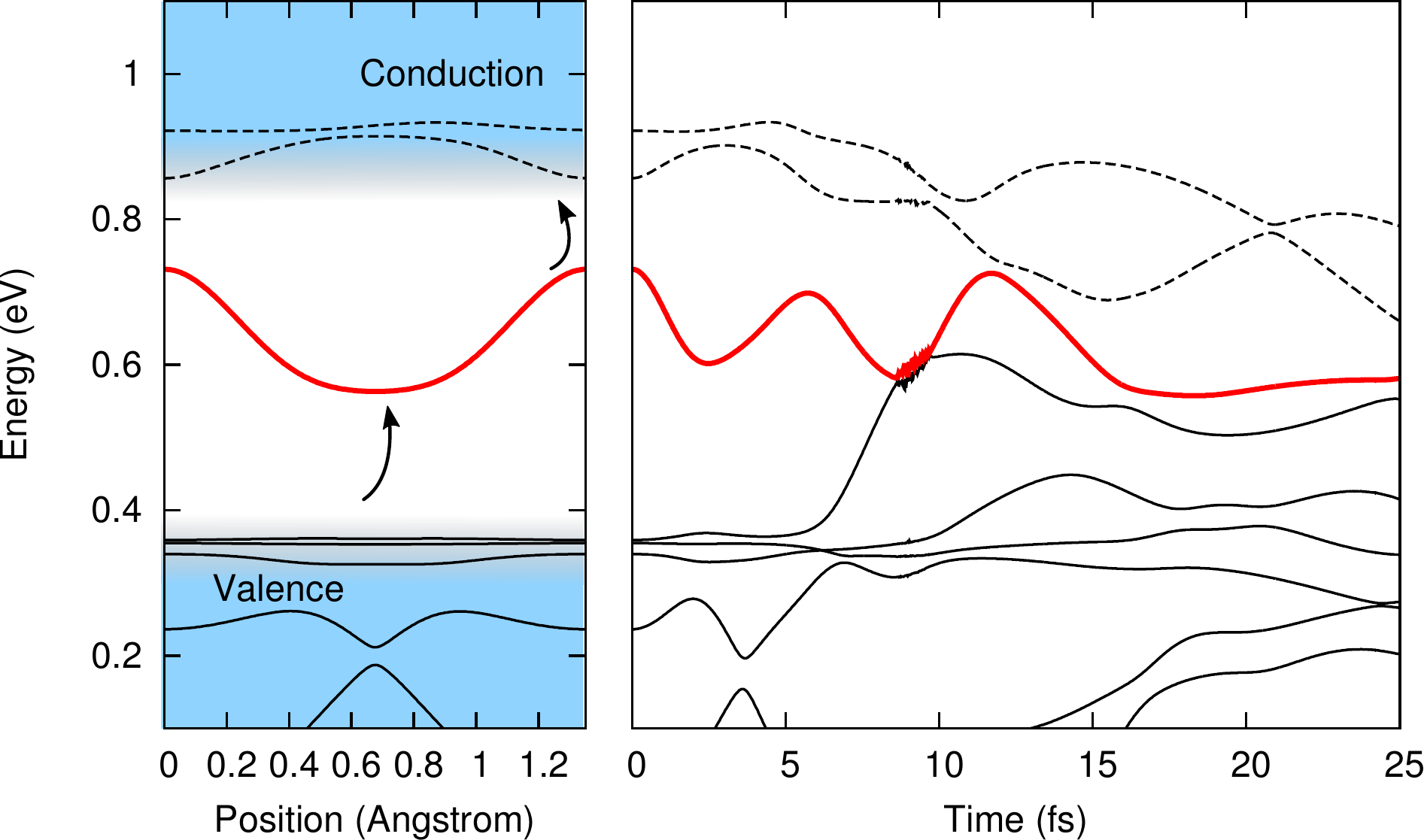}
\caption{
Left: Evolution of the adiabatic energy eigenvalues at the supercell $\Gamma$ point as a function of the position of the projectile along the $\langle 001 \rangle$ channel in Si.
The addition of the channeling ion creates a mid-gap ``elevator state'' (red line), the energy of which oscillates as the projectile moves along the channel.
The smallest difference between the elevator energy and the valence (conduction) band edge is approximately (less than) $0.2\,$eV.
The velocity threshold $E_{\text{th}}'$ calculated using Eq.~(\ref{atompass}) for an energy gap of 0.2 eV is shown in Fig.~\ref{stop_e}.
Right: Evolution of the adiabatic energy eigenvalues during a Born-Oppenheimer quantum molecular dynamics simulation in which a Si interstial in a relaxed crystal of 64 initially stationary Si atoms was given an initial velocity of 0.2 $\text{\AA\,fs}^{-1}$ in the $\langle 001 \rangle$ direction.
All 65 atoms were allowed to move in response to the interatomic forces generated during the simulation.
The initial velocity of 0.2 $\text{\AA\,fs}^{-1}$ is too low for a channeling trajectory to take place, but the elevator-like behavior of the defect states remains apparent.
Similar elevator-like behavior was observed when the initial velocity of the projectile was along the $\langle 110 \rangle$ direction.
\textcolor{red}{}
}\label{track}
\end{figure}

The below-threshold electronic stopping may be understood in terms of the electronic structure of the Si projectile, which acts in many respects like an interstitial.
An adiabatic (Born-Oppenheimer) calculation of the electronic structure shows that the addition of a frozen projectile creates a strongly localized unoccupied defect state near the middle of the gap.
The energy $\epsilon_d(z)$ of the mid-gap defect state $|\psi_d(z)\rangle$ relative to the band edges is shown in the left-hand panel of Fig.~\ref{track} as a function of projectile position $z$.

\begin{figure}
\includegraphics[scale=0.4]{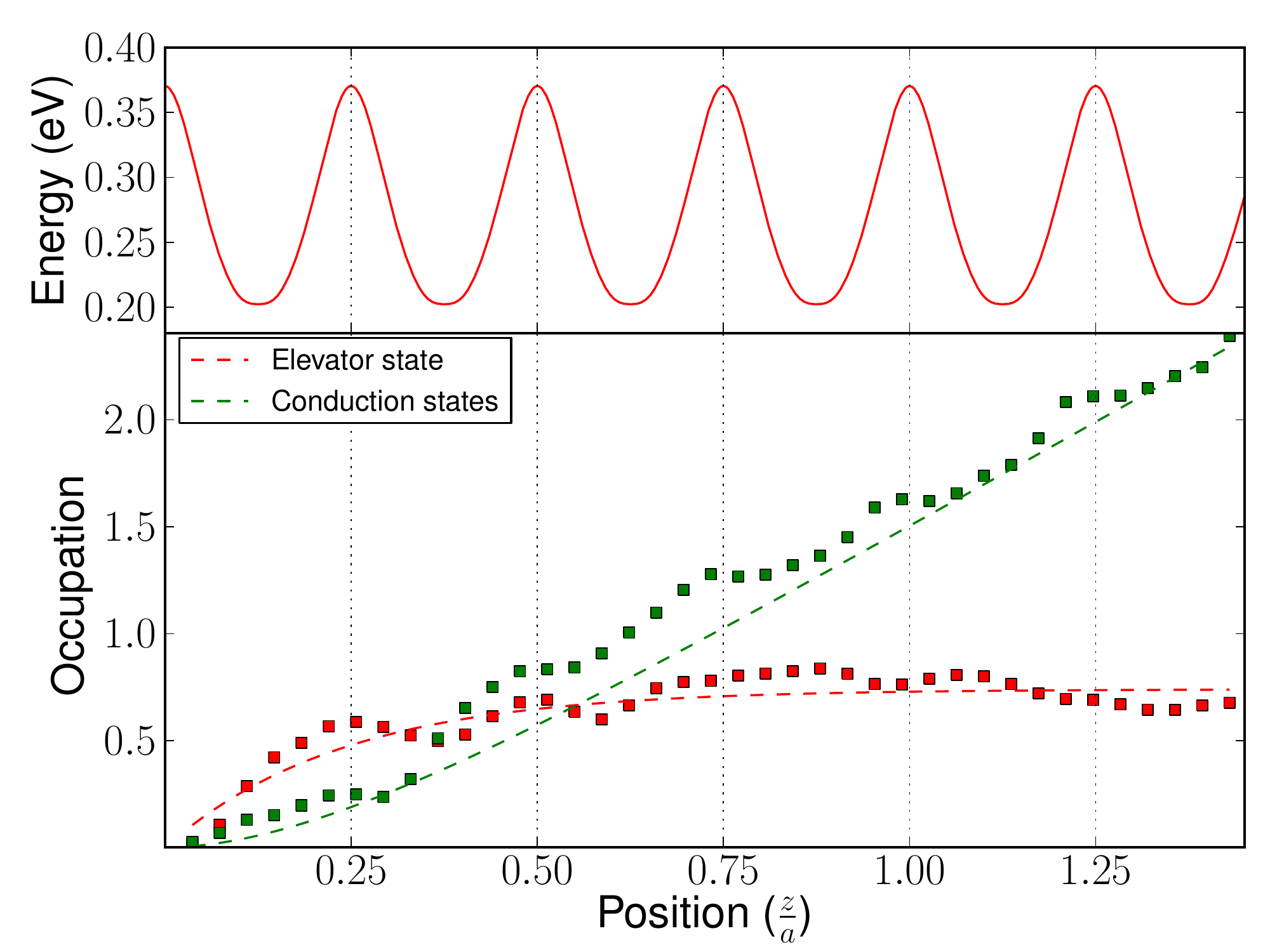}
\caption{
Evolution of the occupation of the elevator state as an ion of kinetic energy $39\,\text{eV}$ ($v \approx 0.164\,\text{\AA\,fs}^{-1}$) travels down the $\langle 001 \rangle$ channel.
The defect state gains electrons when its energy is close to the valence band edge and loses electrons when its energy is close to the conduction band edge; the loss of electrons from the defect state is accompanied by a corresponding increase in the occupation of the conduction band.
The elevator state thus controls the excitations of electrons to the conduction band.
The dashed lines show a fit to the rate-equation model described in the text.
}\label{occ}
\end{figure}

The time-dependent occupation $n_d(t)$ of this state in a non-adiabatic simulation may be calculated by projecting the time-evolved TDDFT Kohn-Sham wavefunctions $|\phi_i(t)\rangle$ on to $|\psi_d(z(t))\rangle$:
\begin{equation} 
n_d(z(t))=\sum_{i\in \text{occ}}|\langle
\phi_i(t)|\psi_d(z(t))\rangle|^2 .  
\end{equation}
Similarly, by summing over conduction band states, we can calculate the population of the conduction bands.
The results in Fig.~\ref{occ} show that, although initially empty, the mid-gap defect level becomes fractionally occupied at the lowest point of its energy oscillation, acquiring electrons from the valence bands via a process analogous to Zener tunneling \cite{zener}.
These electrons are carried along with the defect state until it reaches the highest point of its energy oscillation, at which point the occupation drops as electrons are promoted into the conduction band.
The defect level acts as an ``electron elevator'', ferrying electrons across the gap.
The elevator effect is purely dynamic and is only observed when the projectile velocity is non-zero.
The total electronic energy increases (and the ionic kinetic energy decreases) every time an electron is promoted, either directly across the gap or via the elevator state.

\begin{figure}
\includegraphics[scale=0.4]{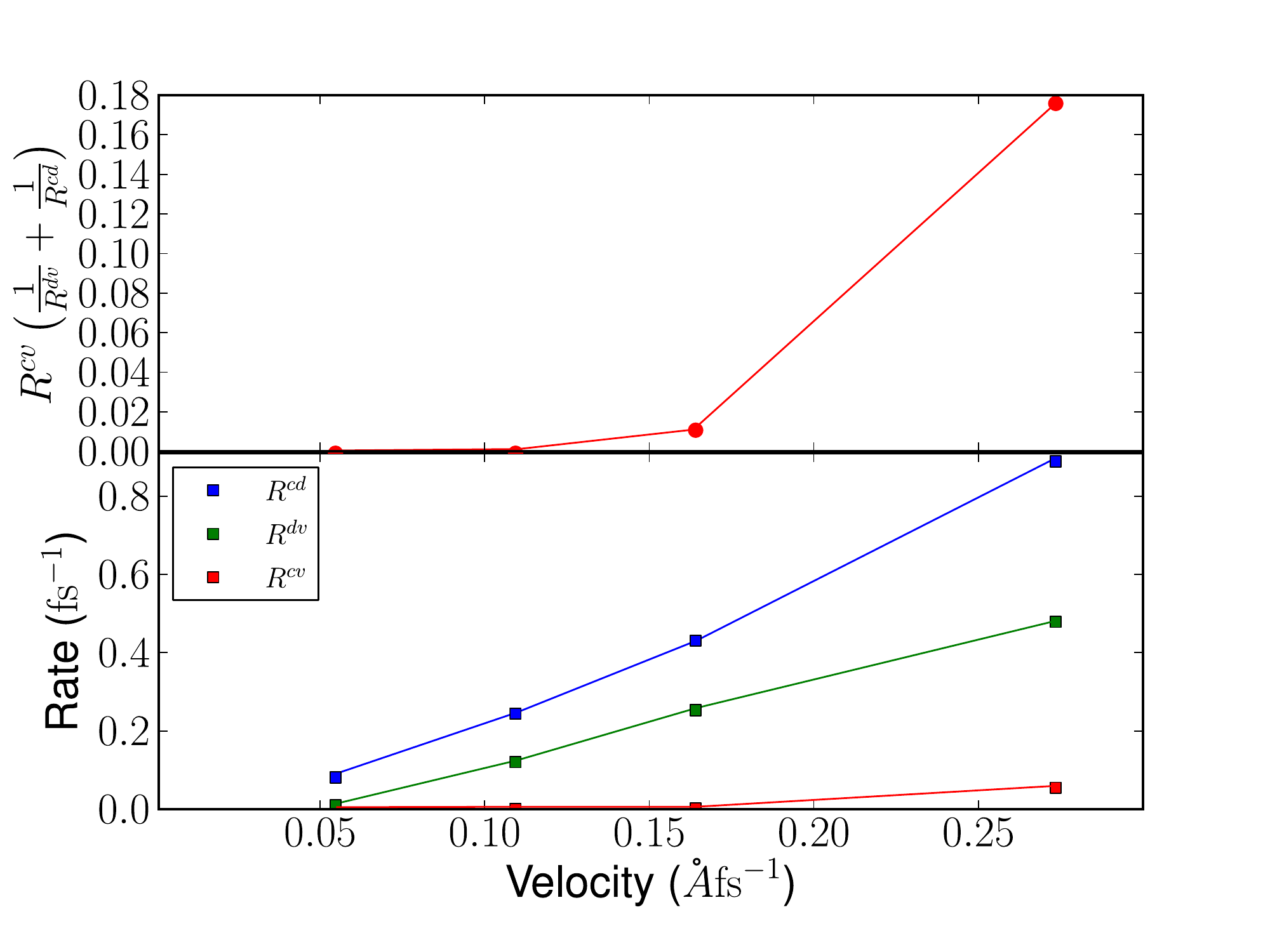}
\caption{
Bottom: the velocity dependence of the rates of excitation directly from the valence band to the conduction band ($R^{cv})$, from the valence band to the defect level ($R^{dv}$), and from the defect level to the conduction band ($R^{cd}$).
Top: the ratio of the rate of direct valence-conduction excitations to the rate of excitations via the defect state. 
}\label{rates}
\end{figure}

Included in Fig.~\ref{occ} are results from a linearized set of coupled rate equations that describe the time evolution of the populations of the states.
(More details are provided in the supplementary material.)
These equations give an accurate description of the time-dependence of the state occupations, as calculated from TDDFT, with just three fitting parameters: the rate $R^{cv}$ at which electrons are excited directly from the valence band to the conduction band, the rate $R^{dv}$ at which electrons are excited from the valence band to the defect level, and the rate $R^{cd}$ at which electrons in the defect level are excited into the conduction band.
The rate equations do not capture the time-dependent oscillations seen in Fig.~\ref{occ} because they assume time-independent states and rate coefficients.
The velocity-dependent rate coefficients are plotted in the lower panel of Fig.~\ref{rates}; the upper panel shows the relative contribution of direct transitions to the total electronic stopping as a function of velocity.
We have also analyzed these rates using adiabatic perturbation theory; the results will be presented elsewhere \cite{my_thesis}.
For this system of one moving Si atom and 216 stationary Si atoms, indirect excitations via the elevator state dominate throughout the pre-threshold regime.

The contribution of the elevator to stopping is most important in the low-velocity below-threshold regime.
Since Si projectiles do not channel through Si at such low velocities, the relevance of the elevator needs clarification.
The right-hand panel of Fig.~\ref{track} shows the evolution of the adiabatic energy levels during a Born-Oppenheimer quantum molecular dynamics simulation in which a tetrahedral Si interstitial in a fully relaxed crystal of 64 stationary Si atoms was given an initial velocity of $0.2\,\text{\AA\,fs}^{-1}$ (corresponding to an initial kinetic energy of 58 eV) in the $\langle 001 \rangle$ direction.
The other atoms, although initially stationary, were all free to move.
Although no channeling takes place, the strong elevator-like behavior of the defect states is apparent.

In summary, direct simulations have shown that there are three regimes in the electronic stopping power of Si in a Si crystal.
In the metallic regime at kinetic energies greater than $3000\,\text{eV}$, the electronic stopping power is directly proportional to the velocity of the channeling ion.
The electronic stopping in this regime is dominated by direct excitations from valence-band states to conduction-band states and is insensitive to the detailed structure of the density of states, including the presence of the band gap.

The band-edge regime is observed for kinetic energies between $60$ and $3000\,\text{eV}$.
The significant increase in the electronic stopping power with kinetic energy observed in this regime originates from the steep rise in the number of states available for direct band-to-band transitions as the kinetic energy increases. 

The pre-threshold regime is observed at projectile kinetic energies below $60\,\text{eV}$, where the electronic stopping power remains non-zero at velocities much lower than predicted by naive estimations of threshold velocities.
The complicated energy dependence of the stopping power in this regime is generated by indirect transitions from the valence band to the conduction band via an elevator level in the band gap.
This suggests that any realistic model of velocity-dependent forces in a material with a band gap should allow for stopping below the thresholds calculated using Eqs.~\ref{atompass} and \ref{di}.
At low velocities the electronic stopping is controlled not only by the band gap but also by the local electronic structure of the defects present in the material.

\begin{acknowledgments}
This work was performed under the auspices of the U.S.\ Department of Energy by Lawrence Livermore National Laboratory under Contract DE-AC52-07NA27344.
Computing support for this work came from the Lawrence Livermore National Laboratory Institutional Computing Grand Challenge program.
Anthony Lim was supported by the CDT in Theory and Simulation of Materials at Imperial College London funded by EPSRC grant EP/G036888/1 and the Thomas Young Centre under grant TYC-101. In part this work has been carried out within the framework of the EUROfusion Consortium and has received funding from the Euratom research and training programme 2014-2018 under grant agreement No.\  633053. The views and opinions expressed herein do not necessarily reflect those of the European Commission.
\end{acknowledgments}

\appendix


\bibliographystyle{ieeetr}

\end{document}